\documentclass[a4paper,times]{cjeart}
\usepackage{subfigure}
\usepackage{amsthm} 

\graphicspath{{figures/}}
\usepackage{algorithm,algorithmic}

\volumeyear{xxxx}
\volumenumber{xx}
\issuenumber{xx}
\journalname{Chinese Journal of Electronics}
\startpage{1}
\DOI{xxx.xxx.xxx.xxx}
\journaltype{Research Article}

\title[CJE Science Paper Template for Submission - Here is the Title of This Article]{Non-stationarity Characteristics in Dynamic Vehicular ISAC Channels at 28 GHz}

\author{%
Zhengyu Zhang\affilnums{1}, 
Ruisi He\affilnums{1}, 
Mi Yang\affilnums{1,2}, 
Xuejian Zhang\affilnums{1}, 
Ziyi Qi\affilnums{1}, 
Hang Mi\affilnums{1}, 
Guiqi Sun\affilnums{1},
Jingya Yang\affilnums{2}, and 
Bo Ai\affilnums{1}
}

\affiliation{%
\affilnum{1}School of Electronic and Information Engineering, Beijing Jiaotong University, Beijing, 100044, China\\
\affilnum{2}Henan High-Speed Railway Operation and Maintenance Engineering Research Center, Zhengzhou, 451460, China \\
}

\corrauth{Ruisi He, Mi Yang}
\email{ruisi.he@bjtu.edu.cn, myang@bjtu.edu.cn}

\authorcopy{Firstname1 Middlename1 Lastname1\emph{et al.}}

\abstract{%
Integrated sensing and communications (ISAC) is a potential technology of 6G, aiming to enable end-to-end information processing ability and native perception capability for future communication systems. As an important part of the ISAC application scenarios, ISAC aided vehicle-to-everything (V2X) can improve the traffic efficiency and safety through intercommunication and synchronous perception. It is necessary to carry out measurement, characterization, and modeling for vehicular ISAC channels as the basic theoretical support for system design. In this paper, dynamic vehicular ISAC channel measurements at 28 GHz are carried out and provide data for the characterization of non-stationarity characteristics. Based on the actual measurements, this paper analyzes the time-varying PDPs, RMSDS and non-stationarity characteristics of front, lower front, left and right perception directions in a complicated V2X scenarios. The research in this paper can enrich the investigation of vehicular ISAC channels and enable the analysis and design of vehicular ISAC systems.
}


\begin{document}

\maketitle

\section{Introduction}
\label{sec1}

With the development of wireless communication systems, the sixth generation (6G) network is considered as the revolution of traditional communication paradigm \cite{A6}. Specifically, it not only represents an enhancement of current communication technologies, but also possesses the capability to provide ubiquitous sensing and connectivity. In this vision, integrated sensing and communication (ISAC) techniques are prompted to provide higher end-to-end information processing ability and native perception ability of wireless devices \cite{1}, which have emerged as a pivotal component in the vehicle industry, playing a crucial role in enhancing vehicle safety and performance. Under the background of deep integration of communication and sensing technologies in the vehicle industry, ISAC-aided vehicle-to-everything (V2X) networks is one of the most potential ISAC application scenarios \cite{A3}. On the one hand, lots of communication transceivers and sensors are equipped on the autonomous vehicles \cite{6179503}. For example, the novel vehicles is usually equipped with eight cameras for a 360-degree of environment perception \cite{10007643}. On the other hand, for ISAC-aided vehicles, it is possible to reuse the current dense deployment road-side units (RSUs) for sensing with only minor modifications in hardware, signaling strategy, and communication standards \cite{A5}. In this way, ISAC-aided V2X networks can connect vehicles with surrounding vehicles, traffic infrastructures, people, and networks, as well as use the integration gain of ISAC to obtain more accurate perception information and better wireless communication quality, so as to reduce traffic jams and accident rates efficiently.

However, ISAC-aided V2X technology is just getting started, and a deep understanding of the radio propagation mechanism of vehicular ISAC channel is essential for its design, performance evaluation and standardization \cite{A4,8,9}. An accurate vehicular ISAC channel reveals the connection between the environment and radio waves, serving as the cornerstone for perception. In general, vehicular ISAC channels do not satisfy the assumption of wide-sense stationarity uncorrelated scattering (WSSUS) due to the high relative mobility of vehicles and scatterers \cite{10,10.5}, and traditional channel models are not applicable in this case. Besides, compared with traditional channel characteristics in V2X scenarios, vehicular ISAC channel is sensitive to the driving environment, such as surrounding roads, vehicles, and other obstacles. Specifically, there are distinct and time-varying environmental features in different perception directions obviously, it is necessary to analyze their channel characteristics accordingly. Therefore, vehicular ISAC channel characterization is a challenging task, the professional channel measurement, characterization, and modeling are imperative for ISAC-aided V2X technology solutions.



Measurement is an effective method to characterize wireless channel \cite{11,11.5}. However, over the past few years, most of the research about ISAC channel measurement and characterization focused static scenarios, lacking studies on dynamic vehicular ISAC channels. Ref \cite{18} conducted sensing channel measurement at 28 GHz and analyzed environment effects. Ref \cite{19} compared channel characteristics between communication channels and sensing channels through actual measurements in outdoor scenarios, and the correlation model of multipaths in ISAC channels are measured and characterized in \cite{20}. Leveraging sensing at the vehicles, the azimuth power spectrum of ISAC channels is measured and analyzed in \cite{21}, which was measured through short-range movement. In \cite{22}, an empirical statistical ISAC channel is presented based on measurements of onboard millimeter wave radar in underground parking lot scenarios, and analyzes multipath components characteristics. Besides, current V2X channel researches are conducted mainly on sub-6 GHz, which is insufficient about millimeter waves.
Ref \cite{12} conducted Vehicle-to Vehicle (V2V) channel measurement at 5.9 GHz in intersection environment and extracted the time-varying PDP and RMS delay spread characteristics of V2V channels. Ref \cite{A1} analyze and model the path loss characteristics for V2V communications based on a series of channel measurements at 5.9 GHz. The underground channel measurements at 2.5/3.5 GHz are conducted in a garage scenario in Ref \cite{A2} to validate the accuracy and usefulness of the proposed vehicular channels. Despite the non-stationarity being one of the main characteristics of V2X channels, existing researches primarily focus on whole periods of channel stability, without taking into account the time-varying changes in perception environment. Ref \cite{16} carried out measurement and analyzed the non-stationarity of V2V and Vehicle-to-infrastructure (V2I) channels in urban scenarios. For more complicated scenarios, non-stationarity vehicular channel characterization are studied in \cite{17}. Although the aforementioned efforts have been conducted to obtain accurate and insightful channel characteristics, dynamic vehicular ISAC channels have not been well measures and characterized.

\begin{figure}[htbp]
\centering
    \includegraphics[width=1\linewidth]{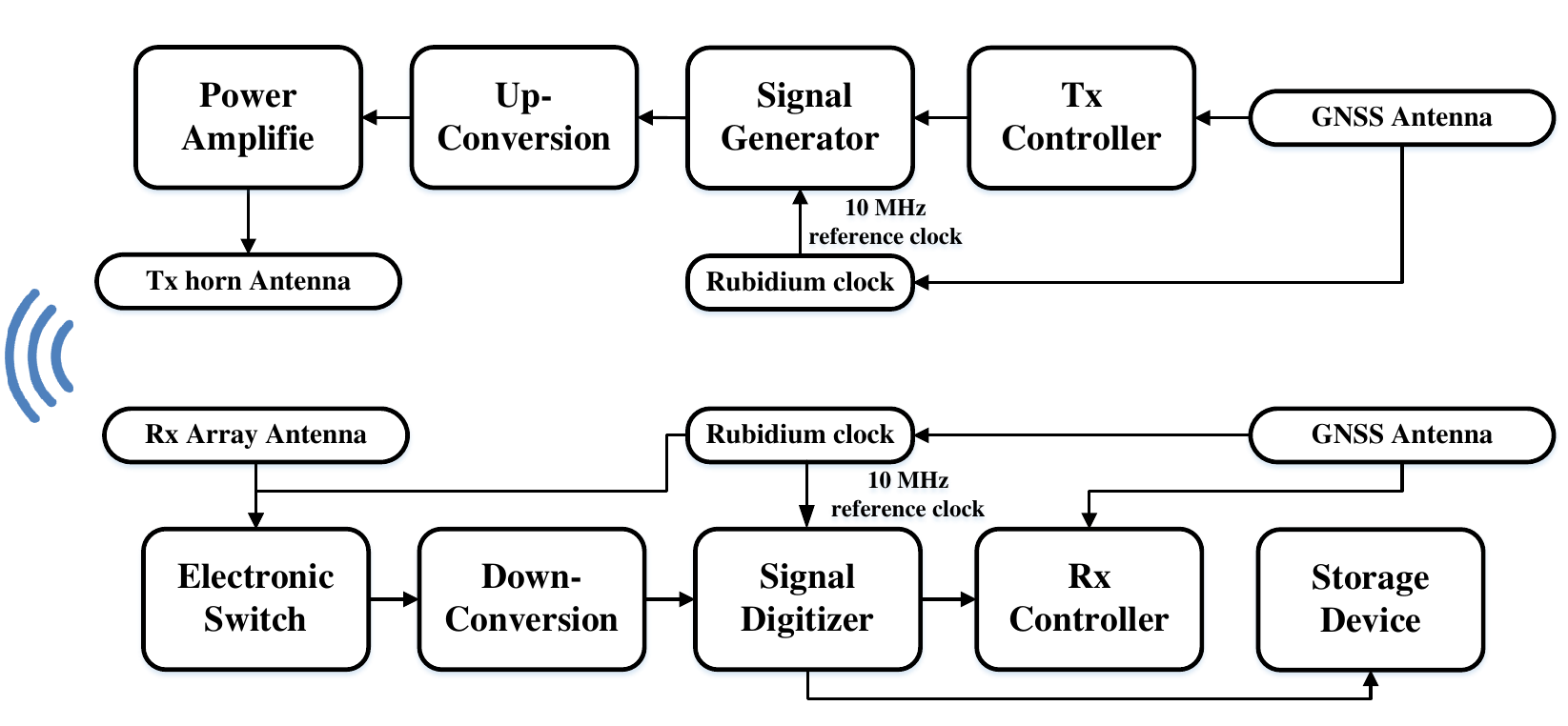}
\caption{The vehicular ISAC channel measurement system.}
\label{fig}
\end{figure}

\begin{figure*}[tbp]
\centering
\subfigure[]{\includegraphics[width=2.9in]{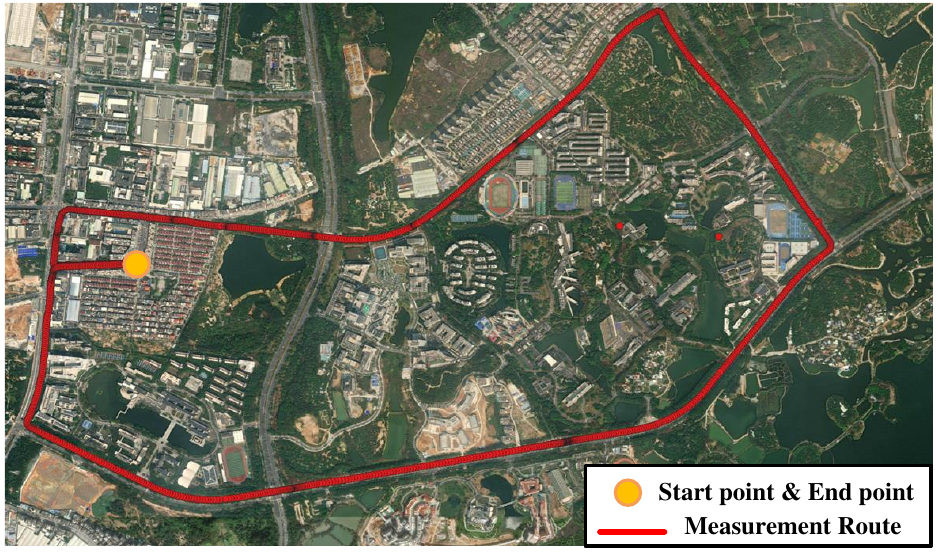}}
\subfigure[]{\includegraphics[width=2.5in]{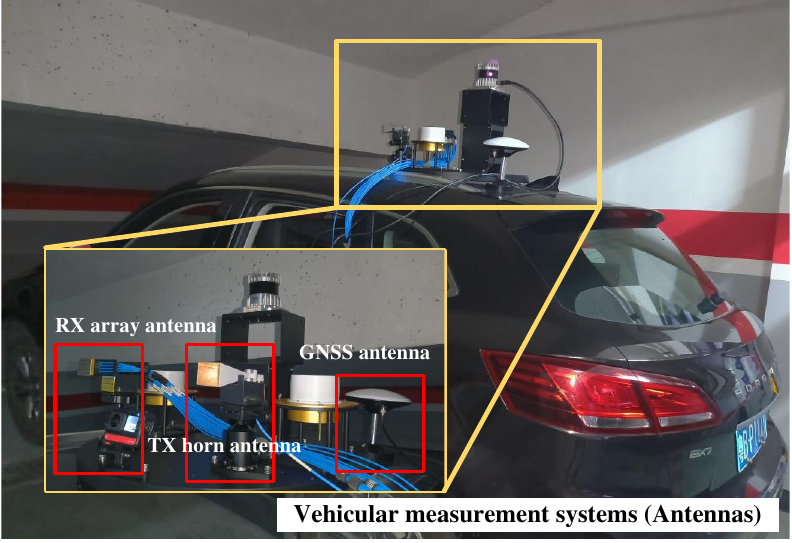}}
\subfigure[]{\includegraphics[width=2.5in]{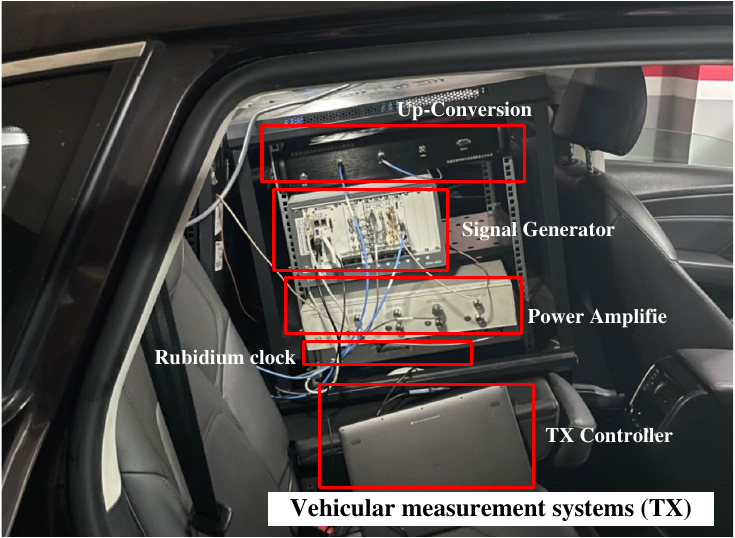}}
\subfigure[]{\includegraphics[width=2.75in]{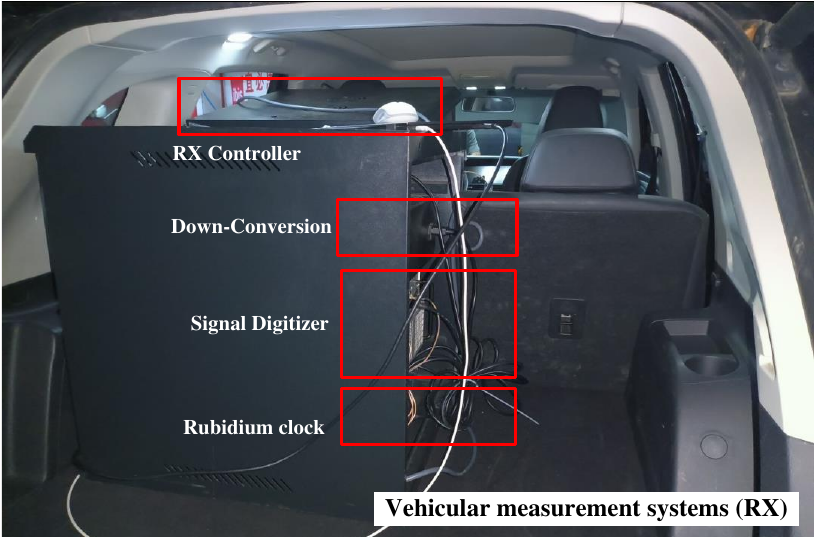}}
\caption{The measurement scenarios and system.}
\label{fig}
\end{figure*}

To fill the gap, ISAC channel measurements at 28 GHz in V2X scenarios are carried out and non-stationarity characteristics are studied. We provide the power delay profile (PDP) and Root Mean Square Delay Spread (RMSDS) to show the time-varying channel characteristics and utilize the temporal PDP correlation coefficient (TPCC) and Spectral Dispersion (SD) to analyze the non-stationarity. Based on this, different perception directions including front, lower front, left and right side are measured and analyzed. The measurement and modeling results demonstrate that the vehicular ISAC channels exhibit significant non-stationarity, with distinct channel characteristics observed for different perception directions. The research in this paper can enrich the investigation of vehicular ISAC channels and enable the analysis and design of ISAC-aided V2X networks.

The remainder of this paper is organized as follows. Section II describes the channel measurement campaigns in details, including the measurement environment and measurement system. Section III and Section IV present and analyze the time-varying PDP, RMSDS and non-stationarity characteristics in dynamic vehicular ISAC channel for different perception directions respectively. Finally, Section V draws the conclusions.

\section{Measurement campaign}

\subsection{Measurement system}

As shown in Fig,1, the vehicular ISAC channel measurement system based on Signal Generator and Digitizer is designed, including measurement system architecture and key equipment. The National Instruments (NI) PXIe-5745 is employed as “signal generator” in TX. The NI PXIe-5775 is employed as “signal digitizer” in RX. At the Tx controller, the transmitter generates the baseband signal by signal generator, and upconvert them to 28 GHz band through the millimeter wave RF transmission module. The measurement signals is sent out by the horn antenna, which has directional beamwidth of 15 degrees, after the power amplifier, and it can reach 28 dBm. At the Rx controller, the array antenna receives the signals, which is then down converted to baseband through the electronic switch. And they are stored by signal digitizer and storage device. The transmitter and receiver are synchronized by the rubidium clocks and GNSS antennas, which can provide 10 MHz reference clock. In our measurement campaign, the measurement signals are a group of sinusoidal signals with different frequency points in the measured bandwidth to obtain a flat power spectrum in measured broadband, which is similar to the orthogonal frequency-division multiplexing (OFDM).

The detailed configurations of the measurement system are presented in Table I. The center frequency is 28 GHz and the measurement bandwidth is 1 GHz, which results in a delay resolution of 1 ns. In this delay resolution, only Multipath Components (MPCs) with propagation distance differences exceeding 0.3 m can be distinguished in the delay domain. The acquisition rate of channel snapshots is 10 snapshots/s when measured. The distance between TX horn antenna and RX array antenna is 0.2 m, and the gain of these antennas are 20 dB and 5 dB respectively. To eliminate the impact of equipment and ensure accurate measurement data, it is necessary to conduct back-to-back measurements prior to the actual measurements. During data processing, the inherent influence of the equipment can be eliminated based on the acquired calibration data.

\begin{table*}[]
\centering
\caption{The configurations of measurement system.}
\begin{tabular}{c|c}
\hline
Configuration              & Values        \\ \hline
Center frequency           & 28 GHz        \\
Bandwidth                  & 1 GHz         \\
Delay resolution           & 1 ns          \\
Transmission power         & 28 dBm        \\
Tx antenna configuration                & Vertical polarized 18 degree horn antenna with 20 dB gain  \\
Rx antenna configuration                & Vertical polarized 32-element rectangular array antenna with 5 dB gain\\
Mean of vehicle speed              & 30 km/h       \\
Tranceiver height          & 2.3 m         \\ \hline
\end{tabular}
\end{table*}
\begin{figure}[tbp]
\centering
\subfigure[]{\includegraphics[width=3in]{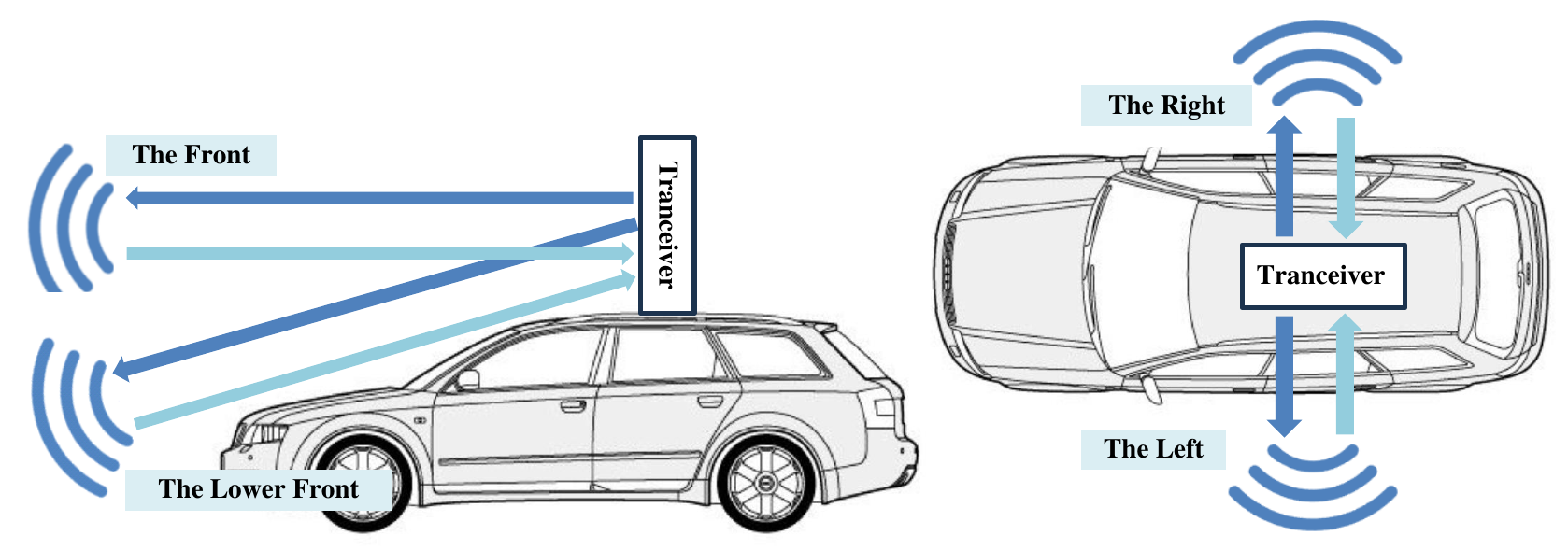}}
\subfigure[]{\includegraphics[width=1.5in]{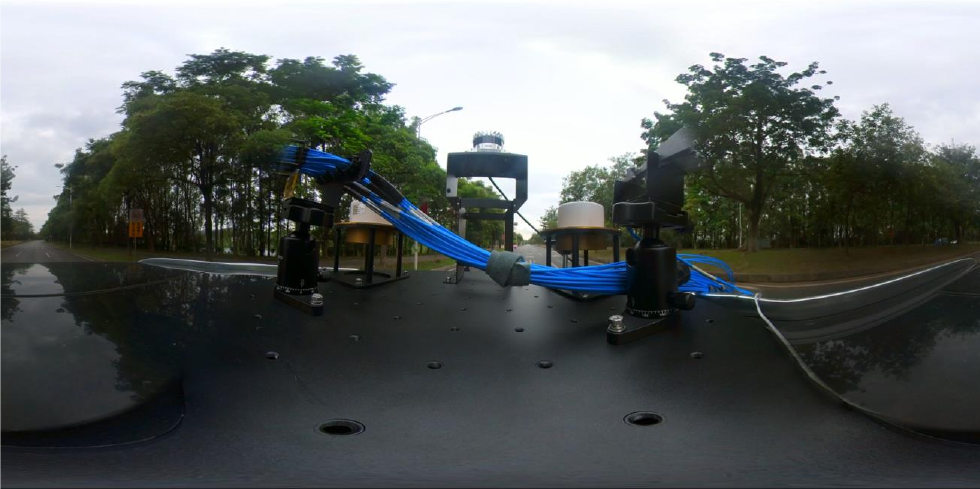}}
\subfigure[]{\includegraphics[width=1.5in]{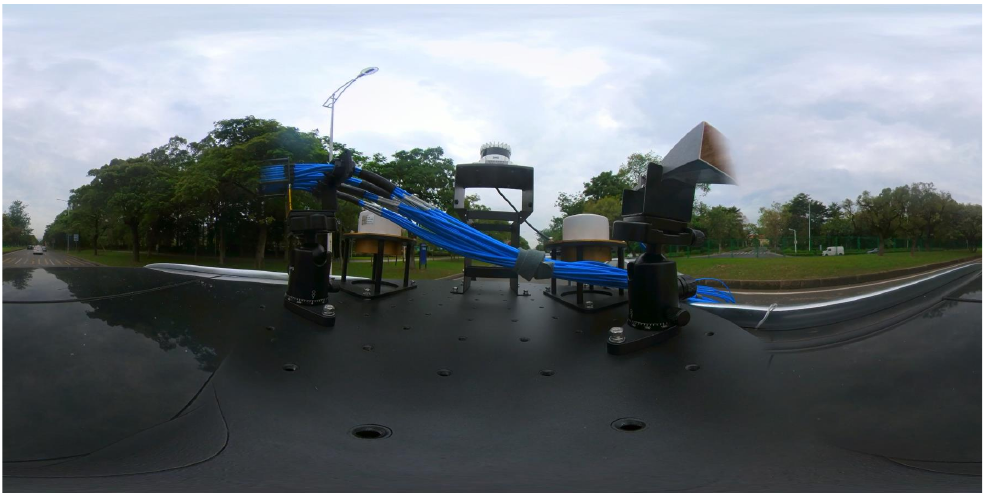}}
\subfigure[]{\includegraphics[width=1.5in]{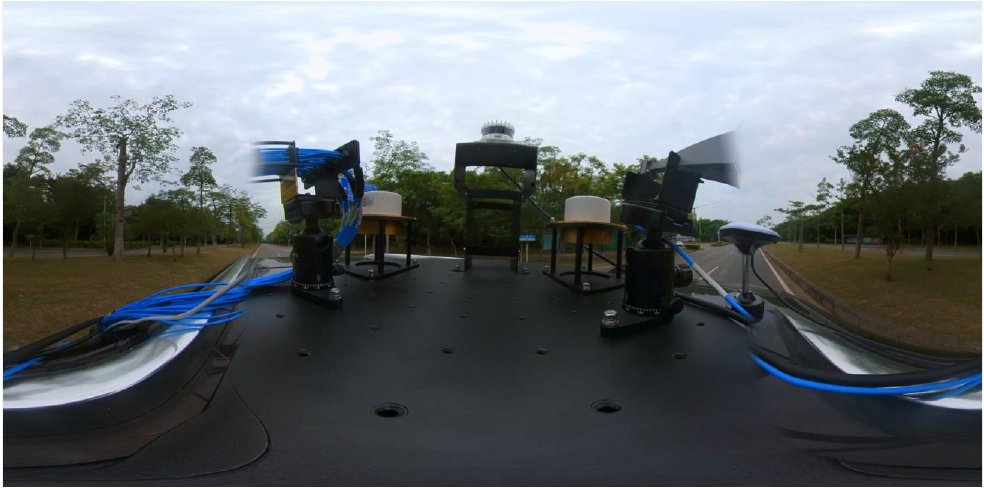}}
\subfigure[]{\includegraphics[width=1.5in]{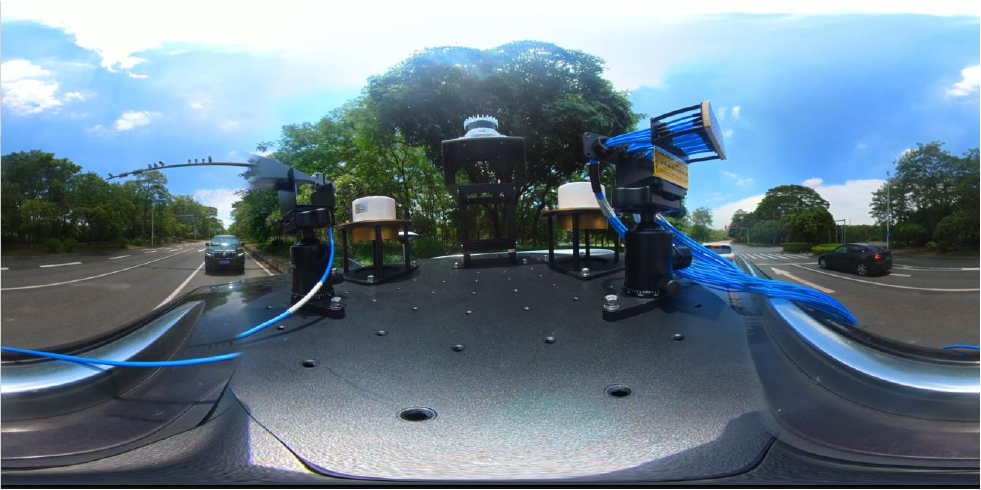}}
\caption{The different perception directions in measurements.}
\label{fig}
\end{figure}
\subsection{Measurement scenarios}

The measurements were conducted in the Songshanhu district of Dongguan, located in Guangdong, China. The measurement-related street is the complicated scenarios, including typical urban scenarios and suburban scenarios. The satellite photographs of the measurement environment are shown in Fig.2 (a). For urban scenarios, the measured streets are surrounded by densely arranged buildings with a height of tens of meters. For suburban scenarios, the measured streets are surrounded by trees and shrubs. The red line in Fig.2 (a) shows the trajectory diagram of the vehicles for a single measurement route. During the measurements, the vehicle’s speed is 0-60 km/h and the duration for measurement cases is generally between 20 and 25 minutes. Lots of buildings, opposite-direction vehicles, parallel direction vehicles, traffic signs, lamp posts, and trees are expected to contribute many sensing MPCs. Fig.2 (b)-(c) illustrates the vehicular measurement systems, which are mounted on vehicles with a height of 2.3 m. The Tx horn antenna, Rx array antenna and GNSS antenna are equipped on the outside of the car, while other equipment is inside the vehicles.

The four measurements are conducted for different perception directions, as shown in Fig.3 (a), including front,lower front, left, and right. The The vision of camera for different perception directions are shown as Fig,3 (b)-(d). Compared to front, lower front has a pitch angle of 30 degrees, making its sensing results more sensitive to vehicles ahead. Different perception directions exhibit varying sensing ranges with diverse scatterers in the environment. For instance, front is more adept at perceiving traffic lights and signs, lower front is better at detecting vehicles ahead, right is more suited for sensing buildings on the roadside, and left is particularly effective for perceiving trees in the middle of the road.

\section{Calculation of channel non-stationarity}

\begin{figure*}[htbp]
\centering
\subfigure[]{\includegraphics[width=3in]{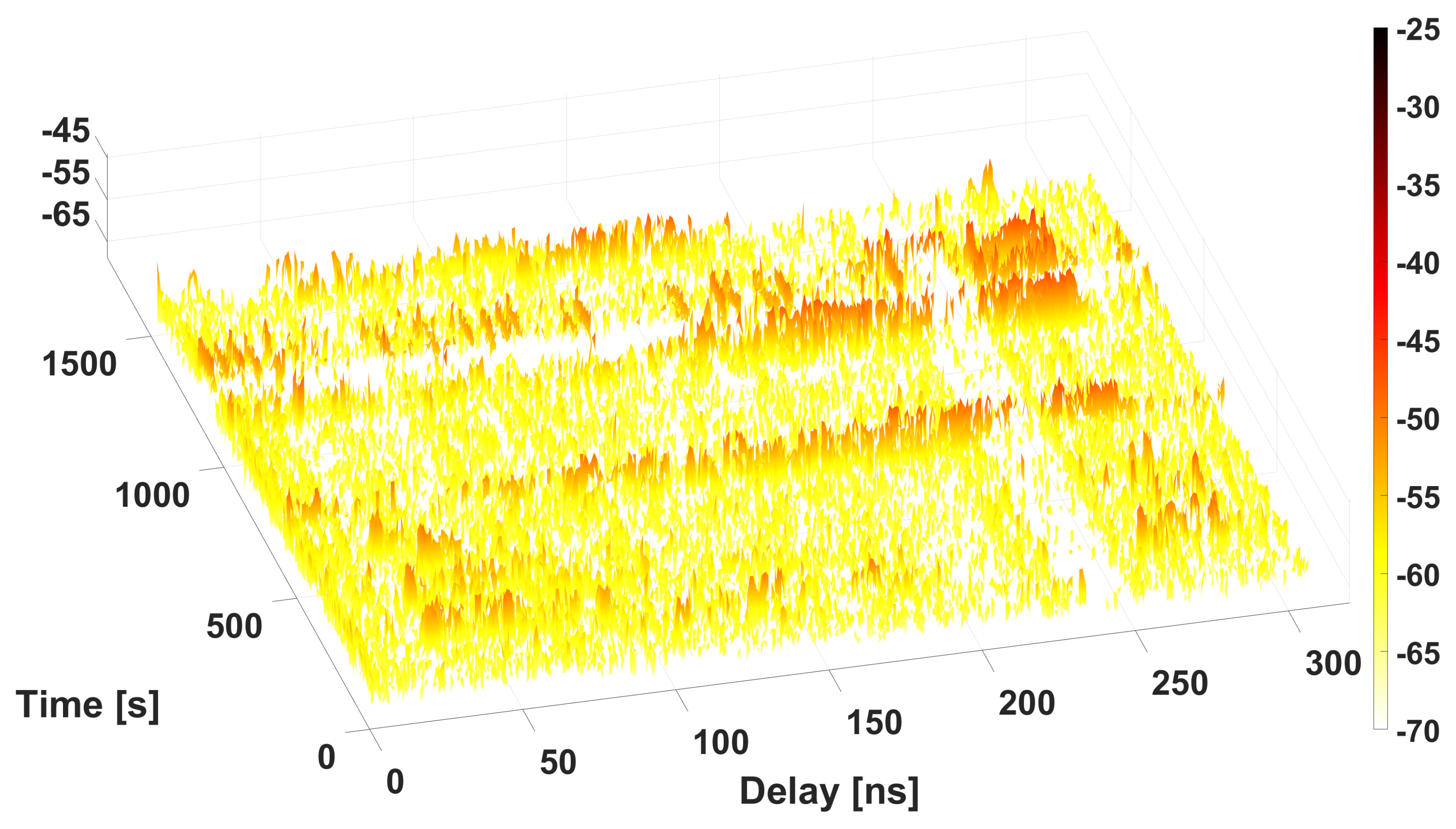}}
\subfigure[]{\includegraphics[width=3in]{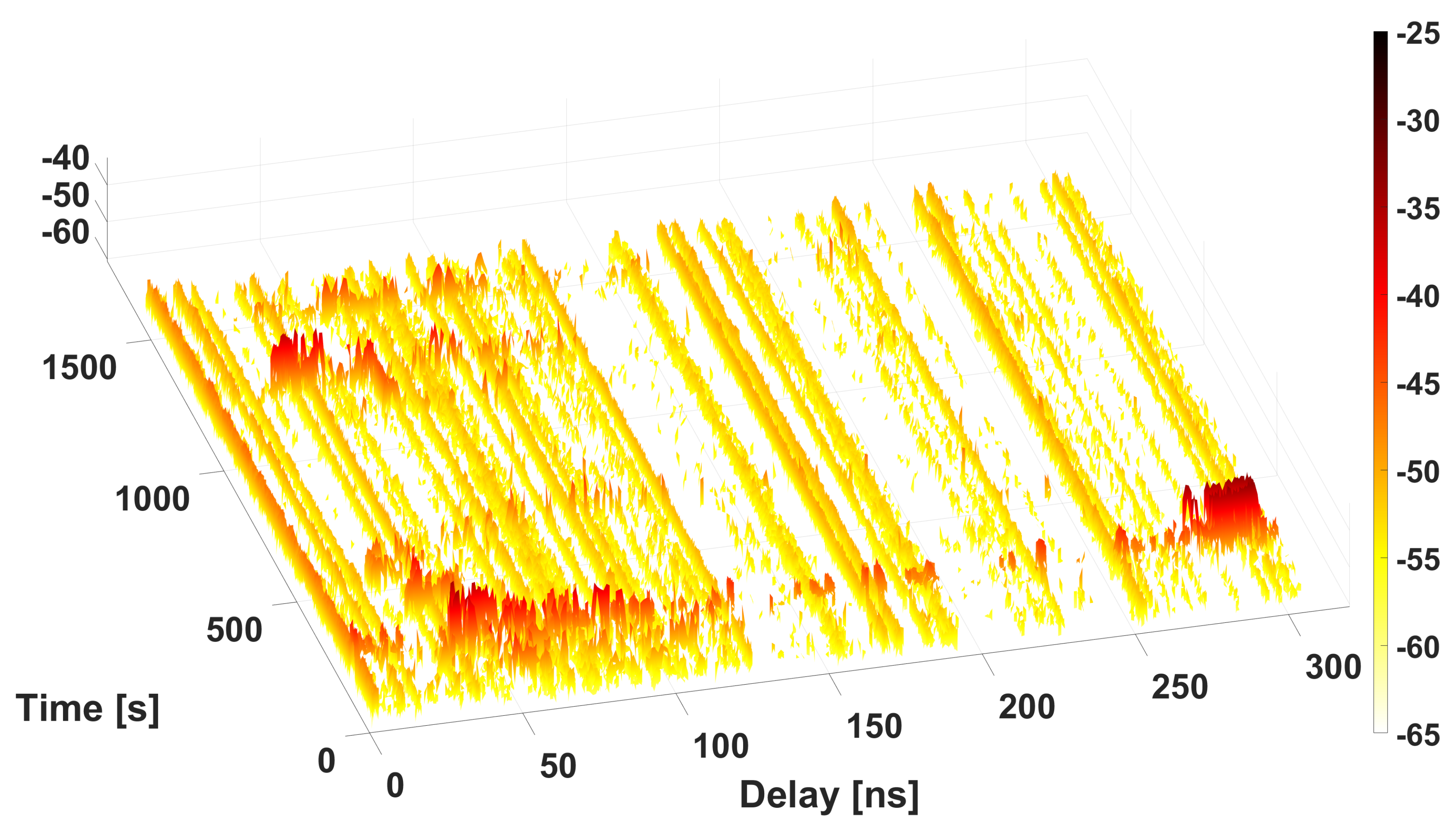}}
\subfigure[]{\includegraphics[width=3in]{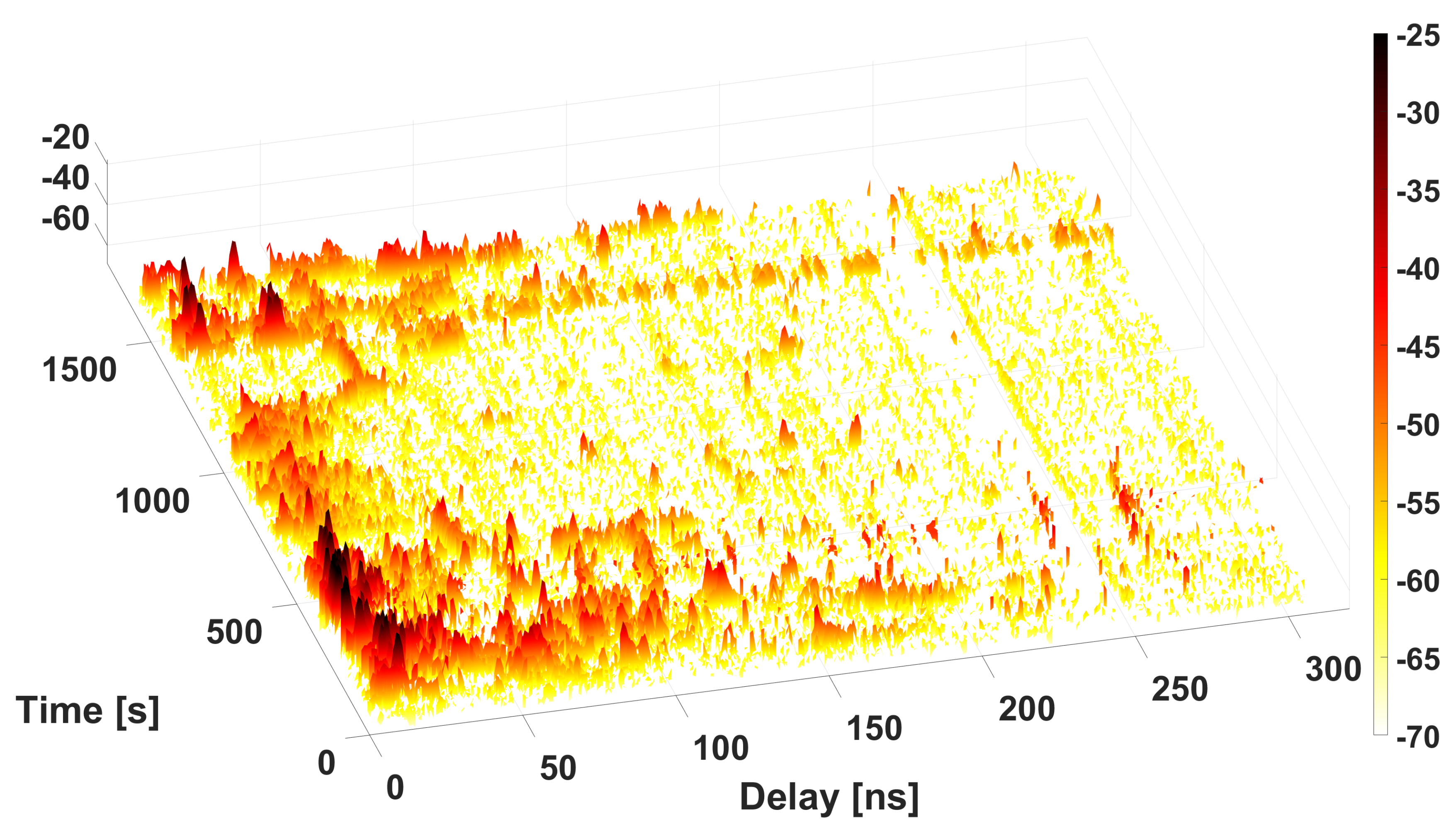}}
\subfigure[]{\includegraphics[width=3in]{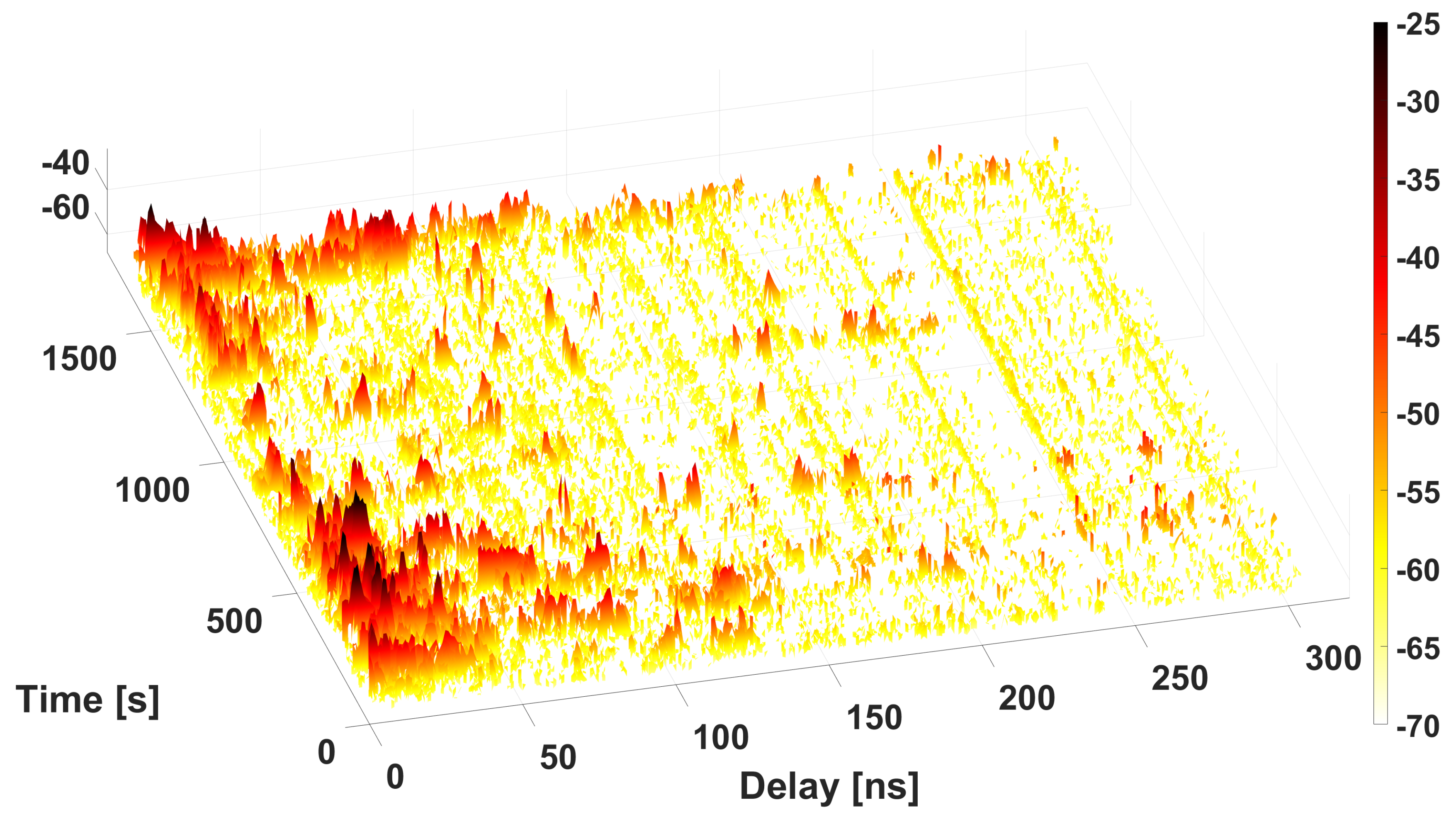}}
\caption{PDPs of vehicular ISAC channels in different perception directions.(a) front. (b) lower front. (c) left. (d) right.}
\label{fig}
\end{figure*}

Due to the movement of the transceiver with the vehicle, causing dynamic changes in the surrounding environment and scatterers, the vehicular ISAC channel is obviously non-stationarity. During short-range areas, the scenarios and distribution of scatterers in the environment are generally similar. However, with the rapid movement of vehicles, the vehicular ISAC channel often undergoes sudden changes in scenarios and scatterer distribution within a short period. In this case, the channel will perform significant time-varying characteristics. The stationarity interval is mathematically the maximum time interval satisfying the wide stationarity condition, which is widely used to characterize the non-stationarity. In this paper, we employ the stationarity interval to quantify the spatial stationarity of channels, so that the statistics in stationarity interval can be approximately regarded as invariant.

To calculate the stationarity interval of vehicular ISAC channels, we first present the channel impulse response (CIR) as follows:
\begin{equation}
h(t, \tau)=\sum_{i=0}^N a_i(t) e^{-j\left[2 \pi f_d(t)\right] t} \delta\left(\tau-\tau_i(t)\right)
\end{equation}
where $t$, $\tau$ and $N$ represent time index, delay and resolvable multipaths number respectively. $a_i(t)$ is the amplitude of the $i$-th multipath, and $f_d(t)$ is the doppler frequency shift.

Power delay profiles (PDPs) are widely used to calculate the received paths with propagation delays, which is obtained by the square of the module of CIR as follows:
\begin{equation}
P D P(t, \tau)=|h(t, \tau)|^2
\end{equation}

The temporal PDP correlation coefficient (TPCC) is used to measure channel stationarity in this paper, and TPCC between PDPs at time $t_i$ and $t_j$ can be calculated as follows:
\begin{equation}
TPCC\left(t_i, t_j\right)=\frac{\int P\left(t_i, \tau\right) \cdot P\left(t_j, \tau\right) d \tau}{\max \left\{\int P^2\left(t_i, \tau\right) d \tau, \int P^2\left(t_j, \tau\right) d \tau\right\}}
\end{equation}

The stationarity interval is defined as the maximum time interval when $TPCC\left(t_i, t_j\right)$ is greater than the given thresholds as follows:
\begin{equation}
T_s(t)=\max \left\{\Delta t \mid T P C C(t, \Delta t) \geq c_{\text {th-TPCC }}\right\}
\end{equation}

Besides, except to TPCC, the spectral dispersion (SD) is also used to measure channel stationarity in this paper. SD metrics are used to describe the difference in PDPs, and a smaller SD suggests a stronger correlation between two snapshots. SD between PDPs at time $t_i$ and $t_j$ can be calculated as follows:
\begin{equation}
S D\left(t_i, t_j\right)=\log _e\left(\frac{1}{N_\tau^2} \int \frac{P\left(t_i, \tau\right)}{P\left(t_j, \tau\right)} d \tau \cdot \int \frac{P\left(t_j, \tau\right)}{P\left(t_i, \tau\right)} d \tau\right)
\end{equation}
where ${N_\tau}$ is the number of delay bins.

The stationarity interval is defined as the maximum time interval when $SD\left(t_i, t_j\right)$ is smaller than the given thresholds as follows:
\begin{equation}
T_s(t)=\max \left\{\Delta t \mid SD(t, \Delta t) \leq c_{\text {th-SD}}\right\}
\end{equation}

In this paper, we select 0.6 and 0.3 as $c_{\text {th-TPCC }}$ and $c_{\text {th-SD }}$ respectively. Specific analysis are shown in Section IV. Besides, a fixed noise threshold is used to eliminate noise components that may be mistaken as multipath, and the noise threshold is set to 6 dB above the background noise floor.

\section{Results and analysis}

\subsection{Power Delay Profile}

As mentioned before, vehicular ISAC channel is sensitive to the driving environment, thus PDPs of different directions perform different dynamic variation.Fig.4 illustrates the PDPs of vehicular ISAC channels in different perception directions, which perform different dynamic channel characteristics. It can be found that the multipath energy in right/left directions is significantly higher than that in front/lower front directions, which is caused by the different perceptual distances and scatterer distributions. While the vehicle is in motion, scattering objects such as trees and buildings along the road will result in stronger echo signals, which makes higher multipath energy in left/right directions. 

Besides, the diversity of multipath in left/right direction is significantly greater than that in front/lower front directions. 
An obvious phenomenon is that front/lower front direction has more high-delay multipath, especially in 100-150 ns, while right/left direction can hardly observe multipath when the propagation delay exceeds 200 ns. In addition, the multipaths in front/lower front direction are sparse in the time domain, and they exhibit a larger range of existence in the delay domain due to the larger spatial depth of perceived targets, such as other moving vehicles. Compared to front/lower front direction, right/left direction can perceive a more diverse set of multipaths due to the chaotic distribution of perceived targets, such as trees. It is noted that multipaths in right/left direction tend to occur at low delays and are uniformly distributed across the entire time domain, and there is no significant difference in channel characteristics between right and left directions.





\subsection{RMS Delay Spread}

\begin{figure}[htbp]
\centering
    \includegraphics[width=1\linewidth]{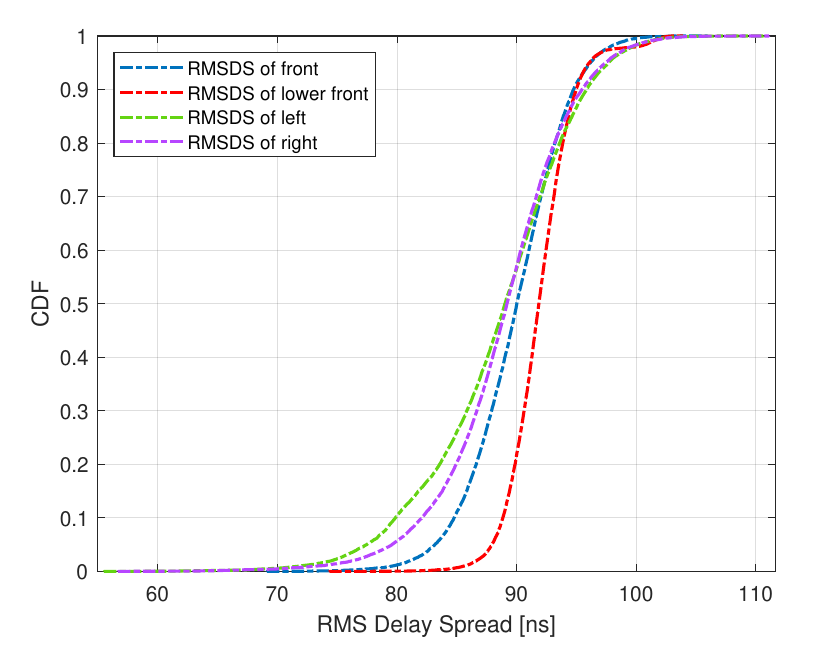}
\caption{The CDFs of RMSDS in vehicular ISAC channels under different perception directions.}
\label{fig}
\end{figure}

The RMSDS is used to characterize the extent of delay dispersion in vehicular ISAC channel.The comparison of RMSDS in different perception directions is presented in Fig. 5. The means of RMSDS in front, lower front, left and right are 89.86 ns, 91.94 ns, 88.44 ns, 88.91 ns respectively, and the standard deviations of RMSDS in the front, lower front, left and right are 4.02 ns, 2.73 ns, 6.21 ns, 5.48 ns respectively. It can be found that there is a significant difference in RMSDS among different perception directions. 

The RMSDS in the front and lower front directions are higher compared to left and right directions. This is because they have larger spatial depth and strong reflectivity of perceived targets, causing these multipaths to still have strong energy even under high delay, resulting in significant delay spread. Although there are strong multipath in left and right directions, their delays are smaller, hence not causing a significant increase in RMSDS. Besides, The standard deviations of RMSDS in left and right are higher compared to front and lower front directions, meaning that there are more diverse scatterers. It is noted that standard deviations of left is higher than right direction. Because on the left side of the road, there are usually diverse scattering objects such as buildings, pedestrians, and trees, while the right side features a median divider for same-direction and opposing traffic. The difference in RMSDS of different directions is not only a reflection of changes in the surrounding environment but also affects the design of ISAC systems.

\subsection{Stationarity Interval}

\begin{figure}[htbp]
\centering
\subfigure[]{\includegraphics[width=1.65in]{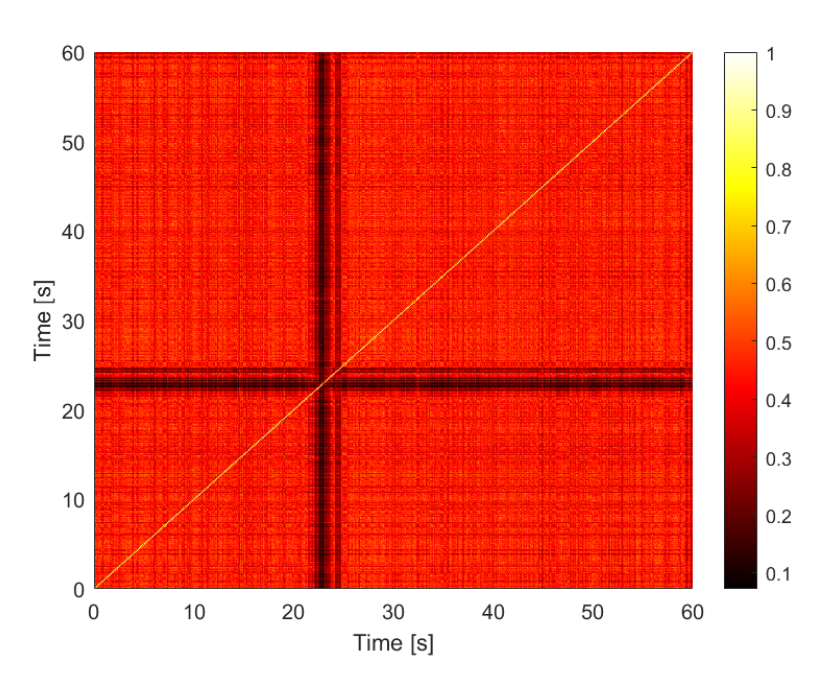}}
\subfigure[]{\includegraphics[width=1.65in]{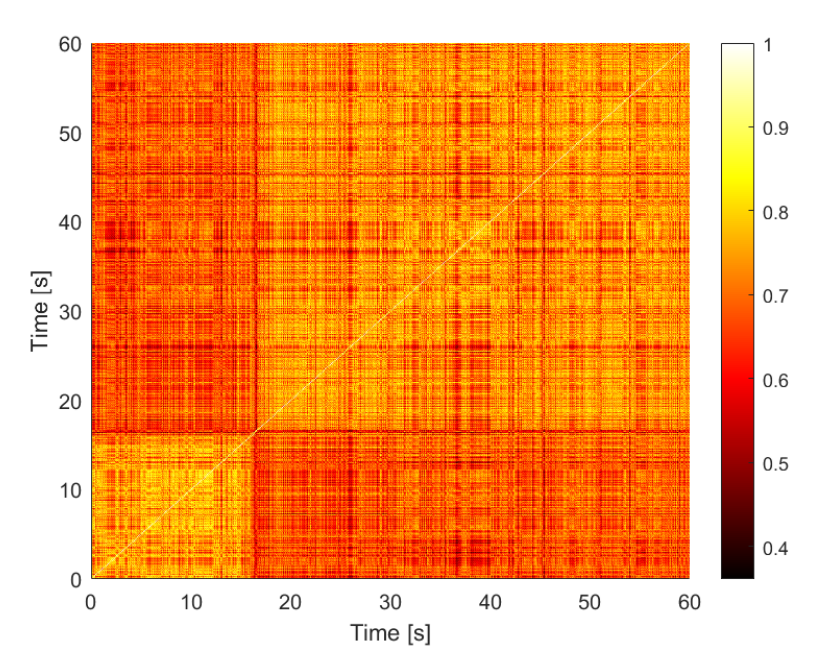}}
\subfigure[]{\includegraphics[width=1.65in]{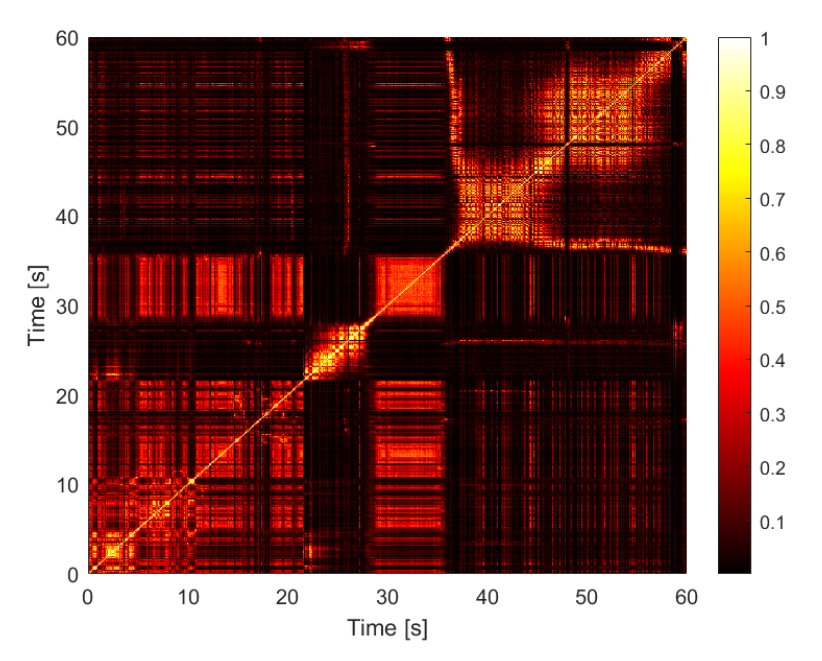}}
\subfigure[]{\includegraphics[width=1.65in]{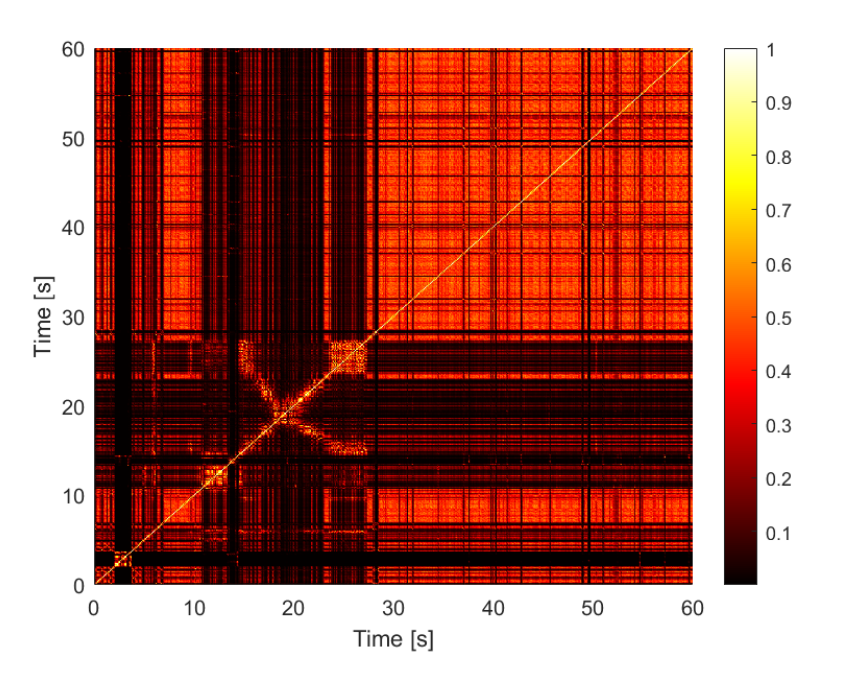}}
\caption{The calculated TPCC matrices of vehicular ISAC channels in different perception directions.(a) front. (b) lower front. (c) left. (d) right.}
\label{fig}
\end{figure}

\begin{figure}[htbp]
\centering
\subfigure[]{\includegraphics[width=1.65in]{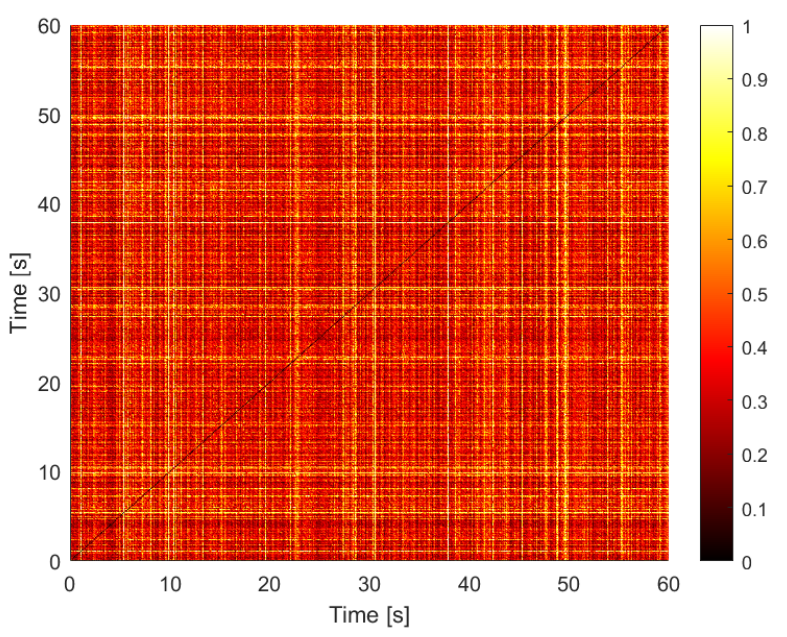}}
\subfigure[]{\includegraphics[width=1.65in]{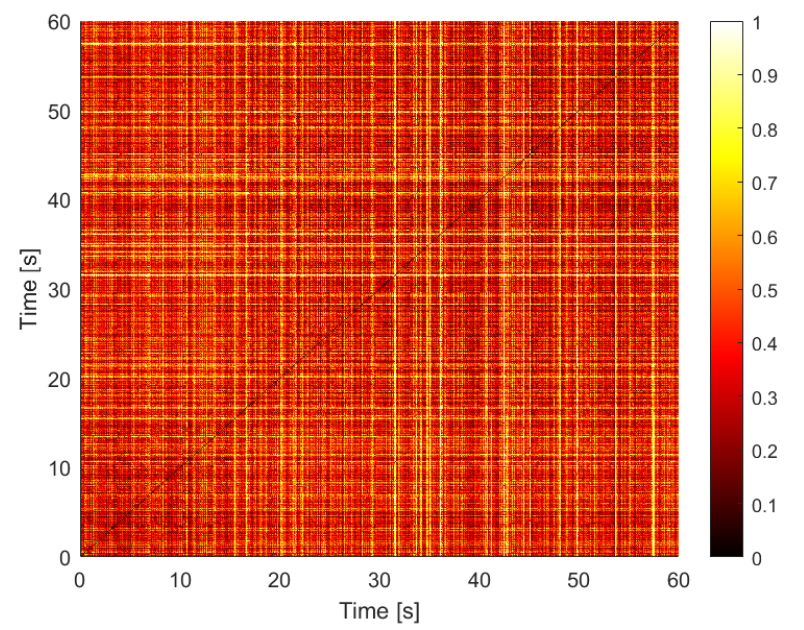}}
\subfigure[]{\includegraphics[width=1.65in]{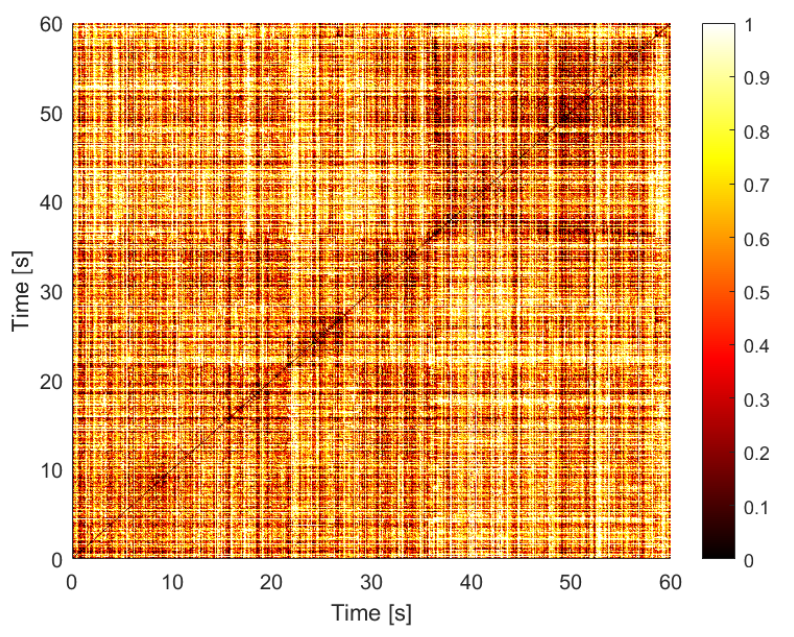}}
\subfigure[]{\includegraphics[width=1.65in]{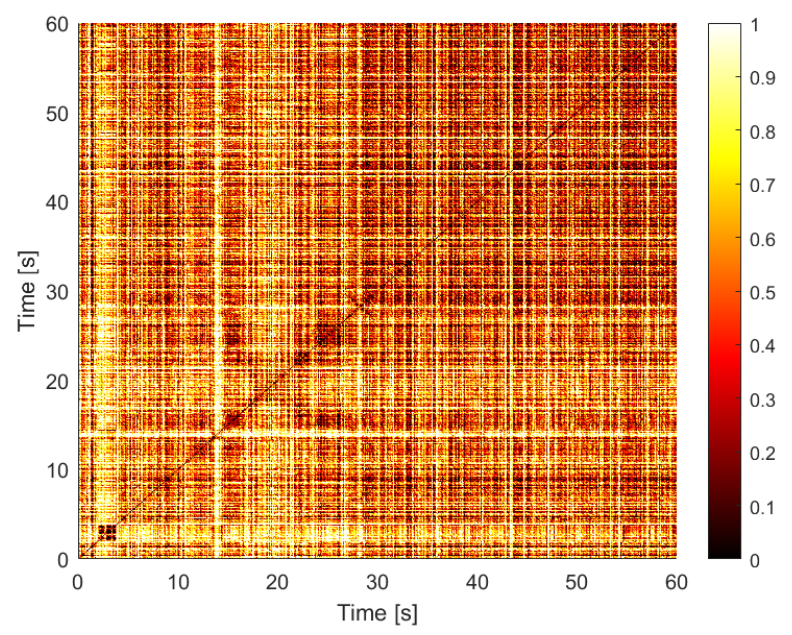}}
\caption{The calculated SD matrices of vehicular ISAC channels in different perception directions.(a) front. (b) lower front. (c) left. (d) right.}
\label{fig}
\end{figure}

\begin{figure}[htbp]
\centering
    \includegraphics[width=1\linewidth]{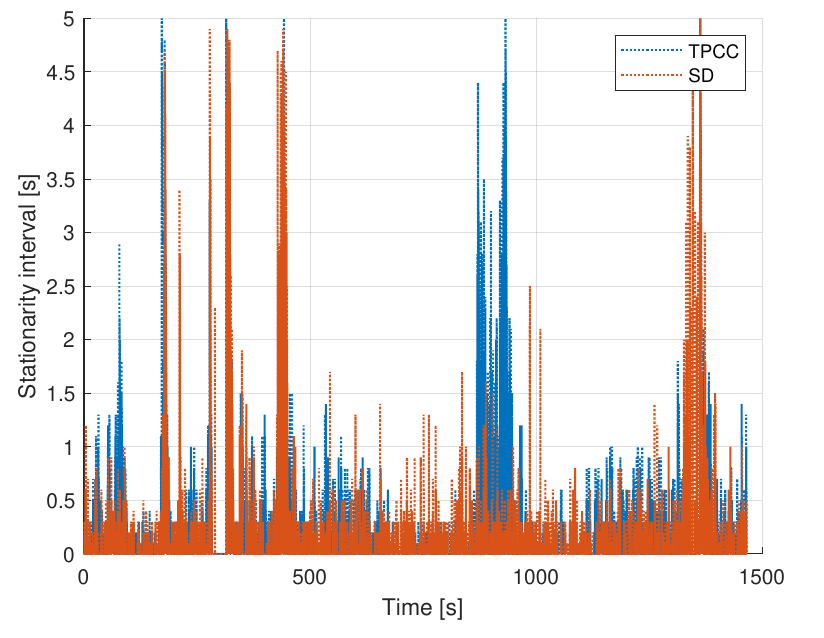}
\caption{The time-varying stationarity intervals for left perception directions with TPCC and SD methods.}
\label{fig}
\end{figure}
In this paper, TPCCs and SDs between each snapshot and all measured samples are calculated for different perception directions, and the matrices are formed. The calculated TPCC matrices and SD matrices (part of all measured snapshots) are illustrated in Fig. 6 and Fig. 7 respectively. Generally, a high TPCC value indicates that the two PDPs have higher similarities, and it is obvious that the diagonal values of the TPCC matrix are 1, and a small SD value indicates that the two PDPs have higher similarities, and it is obvious that the diagonal values of the SD matrix are 0. We can see that with the change of environment, the channel stationarity time is constantly changing, and different perception directions have own characteristics. For front/ lower front direction, the TPCC matrices have higher values (SD matrices have smaller values) because there are usually no clear perception targets in these directions, resulting in more clutter snapshots with higher TPCC values (smaller SD values). It is more obvious in the lower front due to weaker perception echoes than the front. For right/left direction, due to the diverse scatterers and dynamic distribution of scatterers, the TPCC and SD matrices exhibit a more complicated behavior, resulting in more concentrated and smaller stationarity intervals, and they share some similarities.

Fig. 8 shows the time-varying stationarity intervals for left perception directions with TPCC and SD methods. It can be found that there are differences in the calculated results between two methods. For instance, between 800-1000 s, the stationarity interval of TPCC is noticeably larger than that of SD. However, the time-varying trends of stationarity interval for both methods are similar. For instance, there is a significant stationarity intervals between 300-500 s and 1350-1450 s in both methods. It is noted that the majority of stationarity intervals during the entire measurements are relatively small, with larger stationarity intervals concentrated at specific locations. This is because in the high-speed driving, the dynamic changes in surrounding scatterers lead to generally smaller stationarity intervals. However, specific behaviors, such as driving parallel to other vehicles, can cause significant changes in stationarity intervals. Therefore, the stationarity of the vehicular ISAC channel is characterized by the alternating distribution of predominantly small stationarity intervals and behavior-related, sudden large stationarity intervals.

\subsection{Statistics analysis}


\begin{figure}[htbp]
\centering
    \includegraphics[width=1\linewidth]{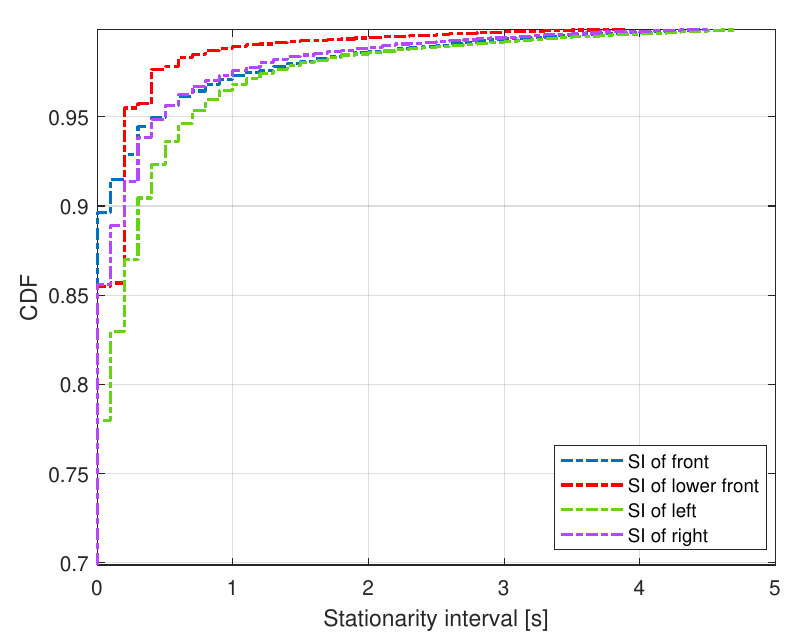}
\caption{The CDFs of stationarity intervals in vehicular ISAC channels via TPCC method.}
\label{fig}
\end{figure}

In order to compare the characteristics of vehicular ISAC channels, we summarize the statistics results of stationarity intervals for different perception directions. Fig. 9 presents the Cumulative Distribution Function (CDF) curves of stationarity intervals under different thresholds via TPCC method, and Fig.10 presents CDFs of stationarity intervals via SD method respectively. For TPCC methods, the stationarity intervals, from largest to smallest, are left, right, front, and lower front perception directions. The corresponding mean stationarity distances are 2.9327 m, 2.0285 m, 1.9911 m, and 1.3609 m respectively. For SD metods, the same order of stationarity intervals is left, right, front, and lower front perception directions, with mean stationarity distances of 3.9691 m, 2.5134 m, 1.8864 m, and 1.4472 m. 

It can be observed that the left perception direction exhibits a more obvious non-stationarity interval, consistently having larger stationarity distances under different methods. The lower front consistently has smaller stationarity intervals under different methods. Besides, the length of stationarity intervals in different directions are generally consistent, with 99$\%$ of stationarity intervals occurring within 2 second. Compared with Ref.[23], in which the stationarity intervals of V2V communication channels are measured as 7.54-10.2 m, the non-stationarity characteristics of vehicular ISAC channels are stronger. Due to the integration of tx and rx, vehicular ISAC channels are more sensitive to driving environmrent, and a stronger non-stationarity implies that the vehicular ISAC channel changes more rapidly than communication channels. The research on non-stationarity contributes to establishing dynamic ISAC channel models, considering multipath effects caused by environments, and designing an ISAC algorithm for an adaptive environment to accommodate different perception cases.

\begin{figure}[htbp]
\centering
    \includegraphics[width=1\linewidth]{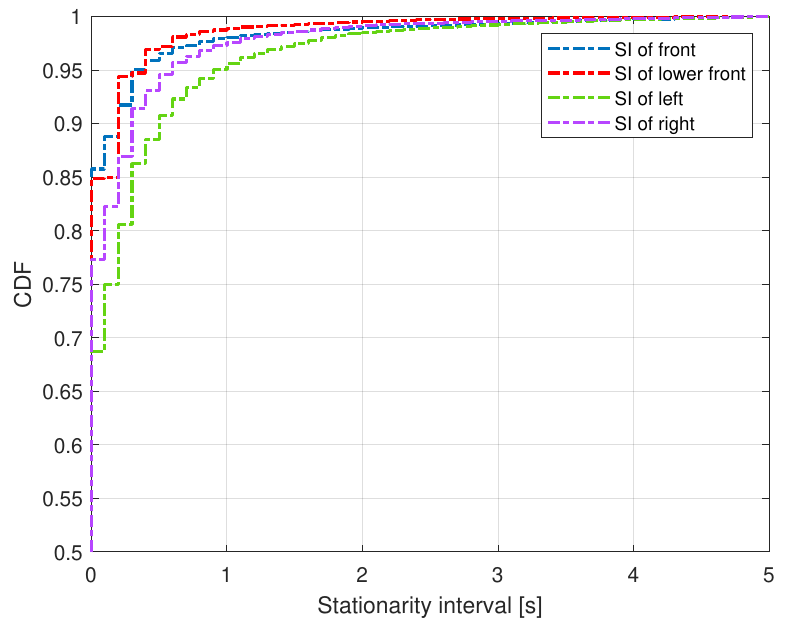}
\caption{The CDFs of stationarity intervals in vehicular ISAC channels via SD method.}
\label{fig}
\end{figure}

\section{Conclusion}

This paper focuses on the dynamic vehicular ISAC channels characteristics for different perception directions. Based on the actual vehicular ISAC channel measurements at 28 GHz, this paper analyzes the time-varying PDPs, RMSDS and non-stationarity characteristics of front, lower front, left and right perception directions in a complicated V2X scenarios. To characterize the non-stationarity, the TPCCs and SDs are adopted to extract stationarity interval in this paper. As a result, the stationarity intervals of vehicular ISAC channels perform alternating distribution of predominantly small stationarity intervals and behavior-related large stationarity intervals. Moreover, we also find that the stationarity intervals of left perception direction are larger than others, which is due to the diverse scatterer’s distribution. The research in this paper can enrich the investigation of vehicular ISAC channels and be the foundation for vehicular ISAC system design and performance evaluation.


\bibliographystyle{cjereport}
\bibliography{refs}

\end{document}